\documentclass[]{aastex631}
\usepackage{subfigure}

\begin{document}

\title{Solar Electron Beam - Langmuir Wave Interactions and How They Modify Solar Electron Beam Spectra: \textit{Solar Orbiter} Observations of a Match Made in the Heliosphere\footnote{Released on XXXX}}

\author[0009-0008-0379-9627]{Camille Y. Lorfing}
\email{camille.lorfing.20@ucl.ac.uk}
\affiliation{Mullard Space Science Laboratory, University College London,
Holmbury St. Mary, Dorking RH5 6NT, UK}

\author[0000-0002-6287-3494]{Hamish A. S. Reid}
\affiliation{Mullard Space Science Laboratory, University College London, 
Holmbury St. Mary, Dorking RH5 6NT, UK}

\author[0000-0002-5705-9236]{Ra\'ul G\'omez-Herrero}
\affiliation{Universidad de Alcalá, Space Research Group,
28805 Alcalá de Henares, Spain}

\author[0000-0001-6172-5062]{Milan Maksimovic}
\affiliation{LESIA, Observatoire de Paris, Université PSL, CNRS, Sorbonne Université, Univ. Paris Diderot, Sorbonne Paris Cité,
5 Place Jules Janssen, 92195 Meudon, France}

\author[0000-0003-3623-4928]{Georgios Nicolaou}
\affiliation{Mullard Space Science Laboratory, University College London,
Holmbury St. Mary, Dorking RH5 6NT, UK}

\author[0000-0002-5982-4667]{Christopher J. Owen}
\affiliation{Mullard Space Science Laboratory, University College London, 
Holmbury St. Mary, Dorking RH5 6NT, UK}

\author[0000-0002-4240-1115]{Javier Rodriguez-Pacheco}
\affiliation{Universidad de Alcalá, Space Research Group,
28805 Alcalá de Henares, Spain}

\author[0000-0001-8661-3825]{Daniel F. Ryan}
\affiliation{University of Applied Sciences and Arts Northwestern Switzerland, Windisch, Switzerland}

\author[0000-0002-0608-8897]{Domenico Trotta}
\affiliation{The Blackett Laboratory, Department of Physics, Imperial College London,
London SW7 2AZ, UK}

\author[0000-0002-0497-1096]{Daniel Verscharen}
\affiliation{Mullard Space Science Laboratory, University College London, 
Holmbury St. Mary, Dorking RH5 6NT, UK}



\begin{abstract}

\textit{Solar Orbiter}'s four \emph{in situ} instruments have recorded numerous energetic electron events at heliocentric distances between 0.5 and 1 au. We analyse energetic electron fluxes, spectra, pitch angle distributions, associated Langmuir waves, and type III solar radio bursts for 3 events to understand what causes modifications in the electron flux and identify the origin and characteristics of features observed in the electron spectrum. We investigate what electron beam properties and solar wind conditions are associated with Langmuir wave growth and spectral breaks in the electron peak flux as a function of energy. We observe velocity dispersion and quasilinear relaxation in the electron flux caused by the resonant wave-particle interactions in the deca-keV range, at the energies at which we observe breaks in the electron spectrum, co-temporal with the local generation of Langmuir waves. We show, via the evolution of the electron flux at the time of the event, that these interactions are responsible for the spectral signatures observed around 10 and 50\,keV, confirming the results of simulations by \citet{Kontar2009}. These signatures are independent of pitch angle scattering. Our findings highlight the importance of using overlapping FOVs when working with data from different sensors. In this work, we exploit observations from all \textit{in situ} instruments to address, for the first time, how the energetic electron flux is modified by the beam-plasma interactions, and results into specific features to appear in the local spectrum. Our results, corroborated with numerical simulations, can be extended to a wider range of heliocentric distances.

\end{abstract}




\section{Introduction}
\label{sec:intro}

The launch of ESA's \textit{Solar Orbiter} in February 2020 has opened a new chapter of solar and space plasma physics \citep{Muller2020}. Among other advances, it paves the way for a better understanding of the transport and kinetics of energetic particles in the heliosphere. 
Energetic electron beams emitted by eruptive solar events travel along magnetic flux tubes through the corona and the solar wind. The interaction with the background plasma they travel through often results in the generation of electrostatic Langmuir waves, with frequencies around the local plasma frequency and its first harmonic, and the subsequent production of solar emission in the radio spectrum known as type III radio bursts \citep[e.g.,][]{Ginzburg, Lin1974,Gurnett1981,Reiner2001,White2011,Reid:2014ab}.

As electron beams propagate away from the Sun, faster electrons outpace the slower ones giving rise to a velocity distribution that shows an enhancement at suprathermal speeds associated with a positive gradient in velocity space. 
 Such a distribution is unstable to a two-stream instability known as the bump-on-tail instability \citep{Ginzburg}, which induces wave-particle interactions that results in the growth of Langmuir waves in the background plasma \citep{Drummond1962,Vedenov1963,Verscharen2022}. The resonant generation of Langmuir waves cause the beam to transfer energy to the background plasma. As consequence of the bump-on-tail instability, the electron distribution function’s shape is modified, with the positive gradient in velocity space flattening into a plateau at energies in the deca-keV range \citep[e.g.][as simulated examples]{Kontar2001a,Kontar2001,Reid2018}. As the Langmuir oscillations are refracted down in velocity space by the density gradient in the solar wind plasma \citep[e.g.][]{Krafft2013}, energy lost by the beam to the plasma is reabsorbed by the lower velocity electrons in the background plasma via Landau damping \citep[e.g.][]{Reid2013,ZZ1970,Zaitsev1972}.

Several spacecrafts (eg. \textit{IMP-6,-7,-8}, \textit{Helios} \citep{Gurnett1976,Gurnett1977}, \textit{STEREO} \citep{Bougeret2008} and \textit{Wind} \citep{Bougeret1995}) have observed electron events, associated Langmuir waves, and type III radio bursts as far out as 1\,au \citep{Thejappa2012,Krafft2016,Vidojevic2017,Thejappa2018}. The peak flux energy spectrum of solar energetic electrons observed in situ typically follows a single or a double power--law \citep{Lin1974,Krucker2009}. Through fitting the electron and X-Ray spectra, the existence of a spectral break in the deca-keV range between 30 and 60\,keV has been found in numerous previous studies \citep[e.g.,][]{Lin1982,Krucker2009,Dresing2023}. Other observations show the existence of a second spectral break above 100 keV \citep{Lin1990,Dresing2021}. More observations of the electron spectrum have identified yet another spectral break at lower energies, around 10 keV \citep{Lin1985,Wang2023}. It is not yet understood why these spectral breaks at different energies happen, or how the electron distribution function is modified \emph{in situ} to support these spectral breaks and other features to appear in the electron spectrum.

Simulations have provided important insights into electron beam parameter space that is not accessible through observations. Solar electron beam transport simulations \citep[e.g.,][]{Magelssen1977,Takakura:1976aa,Li2008,Li2009,Li2011a,Li2011b,Li2012,Li2013,Li2014,Ratcliffe2014,Reid2013,Reid:2015aa,Reid:2017ab,Reid2018} covering the inner heliosphere up to 1\,au, look at the beam-plasma interactions and the modifications of the electron distribution function due to Langmuir wave growth, as well as the modification of the shape of the electron distribution function with the appearance of a plateau at those same energies. Beams of electrons with finite spatial length are able to travel distances up to 1\,au because energy gained at the front of the beam is then reabsorbed at the back, fueling their transport in the heliosphere \citep{Magelssen1977,Takakura:1976aa}.  The efficiency of the Langmuir wave growth can be modulated by the level of fluctuations in the background density gradient \citep[e.g.][]{VoshchKras2015,Reid:2015aa}. Using a one-dimensional Fokker-Planck approach to quasilinear theory, \citet{Kontar2009} simulate the same electron beam both with and without resonant interactions with the background heliosphere plasma. These simulations produce very different results. A beam propagating scatter-free, and not interacting with the solar wind plasma, does not grow Langmuir waves \citep{Kontar2009,Droge2009,Agueda2010} or display any specific features in its electron spectrum assuming that a single power--law is generated at the solar source \citep{Reid2013}. For beams that undergo wave-particle interactions, however, the electron spectrum has a power--law with a spectral break in the deca-keV range \citep{Reid2013} at energies where these beam-plasma interactions occur, in agreement with observations. The broken power law in the electron spectrum forms because of this energy loss. A positive density gradient however, refracts Langmuir waves to higher phase velocities, causing them to be reabsorbed by higher energy electrons \citep[e.g.,][]{Reid2013,Voshch2015,VoshchKras2015}. The spectral index below the break energy depends upon the initial conditions of the electron beam, the distance from the Sun \citep{Reid2013} and the level of density turbulence in the background plasma \citep{Reid:2010aa}.

So far, no \textit{in situ} observational work has looked into the modification of the velocity distribution function, or what electron energies are associated with the growth of Langmuir waves at different distances from the Sun due to limitations caused by the low temporal resolution of sensors onboard spacecrafts like \textit{Helios}, \textit{STEREO}, or \textit{Wind} \citep{Lin1985}. This is now made possible by the high temporal and spectral resolution of instruments onboard \textit{Solar Orbiter}. Previous simulations of wave-particle interactions \citep{Lorfing2023} show that the maximum beam electron velocity interacting with the Langmuir waves decreases as a function of distance from the Sun. The exact maximum electron velocity that interacts with Langmuir waves depends upon the initial electron beam parameters such as beam density and energy spectrum. We therefore still expect to detect Langmuir waves locally, or its signature type III radio emission associated with electron events measured by \textit{Solar Orbiter}.

In this work, we use \textit{in situ} observations in the inner heliosphere to address, for the first time, how the energetic electron distribution function is modified by its interaction with the background plasma, and how this translates into specific features of the local spectrum. Our results exploit data from all four \textit{in situ} instruments of \textit{Solar Orbiter} and are corroborated with numerical simulations.\\

\section{Observational Data}

We use data from the four \textit{in situ} instruments on board \textit{Solar Orbiter}, namely the Energetic Particle Detector~\citep[EPD,][]{EPD}, the Radio Plasma Waves instrument~\citep[RPW,][]{RPW}, the Solar Wind Analyser~\citep[SWA,][]{SWA}, and the Magnetometer \citep[MAG,][]{MAG}. The data is publicly available at \href{http://soar.esac.esa.int/soar/}{Solar Orbiter Archive (esa)}.

\subsection{The Energetic Particle Detector (EPD)}
\label{EPD}

\textit{Solar Orbiter} EPD \citep{EPD} focuses on three main scientific goals: the injection, the acceleration mechanisms, and the transport of solar energetic particles. To study these phenomena, EPD measures amongst other metrics, the distribution function of solar electrons with a maximum time resolution of 1s, through its SupraThermal Electrons and Protons (STEP) (2-80 keV), Electron Proton Telescope (EPT) (25-475 keV) and High Energy Telescope (HET) (450\,keV-18.8\,MeV) units. The STEP detector is mounted on the spacecraft such that its field-of-view (FOV) is centred around the nominal Parker spiral direction (28$^{\circ}$ x 54$^{\circ}$), and particles are collected by 15 sectors (pixels) ~\citep[see][for further info]{EPD}. EPT measures electrons and protons (ions) by using the magnet/foil technique that has been sucessfully used in the SEPT instrument onboard the STEREO mission. There are two EPT units each with two FOVs: EPT1 is pointing sunward and anti-sunward along the nominal Parker spiral at 0.3 au and EPT2 that points northward and southward of the ecliptic plane. Each EPT FOV has an apperture of 30$^{\circ}$. The structure for STEP level 2 data underwent some changes in October 2021. Prior to that date, data is binned into 48 energy channels for the average electron flux and 8 channels for the pixelwise info, while after that date electron flux information is available in 32 energy channels for both average and pixelwise info. The time resolution is 1s in the latter case and 10s in the former data structure.

Pitch angle distributions of energetic electrons are used for the events presented in this work. EPD-STEP is characterised by a relatively narrow FOV around the nominal Parker spiral, but the 15 sectors yield high resolution in pitch angle for the FOV covered by the sensor. In order to compute the pitch angle distributions, EPD-STEP pixelwise data is combined with the magnetic field measurements obtained using the normal mode of the \textit{Solar Orbiter} flux-gate magnetometer MAG, available at a resolution of 8 magnetic field vectors per second. The pitch angle associated with of each pixel is determined using the local magnetic field measurements, with fluxes for overlapping pixels in pitch angle space averaged~\citep[see][]{WimmerSchweingruber2020}.

\subsection{The Radio Plasma Waves instrument (RPW)}

\label{RPW}

The Radio Plasma Waves instrument (RPW) \citep{RPW} measures \textit{in situ} magnetic and electric fields, and plasma waves over a frequency range from almost DC up to a few hundreds of kHz, as well as solar radio emissions up to 16 MHz. It is comprised of several subsystems, two of which are used in this study: the Thermal Noise Receiver (TNR) and the biasing unit (BIAS). The TNR provides the voltage power spectral density used to produce radio dynamic spectra on which Langmuir waves are also observed. Langmuir waves can be measured by RPW using both the Time Domain Sampler (TDS, see  \citep{RPW} ) but also the TNR. While the TDS directly measures the waveforms of Langmuir waves, the TNR measures the total integrated power of the waves. This last quantity is sufficient for the analysis that we have to carry out for this study.
In addition we use the electron density defined from the spacecraft potential and measured by the BIAS unit \citep{Khotyaintsev2021}. 


\subsection{The Solar Wind Analyser (SWA)}
\label{SWA}

The Solar Wind Analyser \citep[SWA;][]{SWA} suite measures the solar wind thermal and suprathermal charged particle populations \textit{in situ} through its three sensors. This study uses data from the Electron Analyser System (SWA-EAS), designed to measure solar wind electrons and resolve their three-dimensional velocity distribution functions. The instrument comprises two electrostatic analyser heads with aperture deflectors and multi-channel plate detectors. Each head measures electrons in the energy range between $\sim$1\,eV and 5\,keV and has a field of view covering $\sim$90$^{\circ}$ in elevation direction and 360$^{\circ}$ in the azimuth direction. In order to have a full-sky coverage, except for small blockages by the spacecraft and its appendages, the two EAS heads are orthogonally mounted at the end of a long boom, extending in the spacecraft shadow. The energy range is resolved in 64 steps by applying discrete voltages on the electrostatic analyser, while the elevation angle is resolved in 16 electrostatic scans of the aperture deflector. For each acquisition in each energy-elevation combination, the azimuth direction is resolved simultaneously by 32 anodes installed on the position sensitive detector (Multi-Channel-Plate, MCP).

\subsection{The Magnetometer (MAG)}
\label{MAG}

The Magnetometer continuously measures the local magnetic field at the spacecraft. The instrument is comprised of two fluxgate sensors, MAG-IBS and MAG-OBS, mounted at 1\,m and 3\,m respectively the boom of the spacecraft. This study uses level 2 magnetic field data in RTN coordinates for a $\pm$10$^{10}$\,nT range. The high temporal resolution of its measurements track variations in scale of the B field. 

Electrons pitch angle distributions have been obtained using MAG and STEP data. To this end, MAG data is downsampled to the STEP time resolution. Then, the look directions of each pixel for each energy bin are combined with magnetic field vectors to obtain the electron pitch angles. For pixels with overlapping pitch-angle coverage, an average flux is computed (as done in ~\citet{WimmerSchweingruber2020, Yang2023}). No Compton-Getting correction~\citep{ComptonGetting1935} was performed, important when looking at ion pitch-angles, due to the lower particle velocities involved in that case~\citep[see][for example]{Yang2020}.

\subsection{EUV Imagers: EUI \& SDO/AIA}

To observe the structure of the solar atmosphere where our electron events originate, we use extreme ultraviolet (EUV) observations from the Atmospheric Imaging Assembly (AIA) onboard the \textit{Solar Dynamics Observatory} (SDO) \citep{Lemen:2012aa}, and the Extreme Ultraviolet Imager (EUI) onboard \textit{Solar Orbiter} \citep{EUI}. Both instruments use bandpass filters to image the thermal emission emitted by coronal plasma around 17.1 nm and 17.4 nm, respectively, corresponding roughly to 1 MK plasma.
Plasma at these temperatures resides in the corona and, due to the low plasma beta there, is frozen onto the magnetic field.
Observations in these passbands therefore give a good indication of the coronal magnetic geometry, and hence the locations from where escaping electrons beams may emanate.

\subsection{The Spectrometer/Telescope for Imaging X-rays (STIX)}

The Spectrometer/Telescope for Imaging X-Rays \citep[STIX;][]{STIX} onboard \textit{Solar Orbiter} provides imaging spectroscopy in the hard X-ray (HXR) regime from 4\,--\,150\,keV.
Due to the difficulty in focusing such high energy X-rays, STIX employs an indirect imaging technique.
Spatial information is encoded into moiré patterns by pairs of absorbing grids and measured by pixellated spectroscopic X-ray detectors.
Images are then reconstructed via algorithms similar to those used in radio astronomy.
While lower energy emission ($\lesssim$\,15\,keV) tends to be dominated by thermalised electrons, higher energy emission ($\gtrsim$\,20\,keV) tends to be dominated by non-thermal electrons impacting the chromosphere.
These electrons are thought to be originally accelerated in the corona via processes associated with magnetic energy release.
STIX therefore reveals the spatial, temporal and spectral evolution of accelerated electrons along closed fields lines as well as the solar atmosphere's thermodynamic response.
Escaping electron beams are thought to originate from the same or neighbouring locations where the magnetic reconnection has created open, rather than closed, field lines.
Hence STIX is another useful aid in determining the most likely locations from which escaping electron beams may emanate.


\section{Event selection}
\label{select}

\begin{figure*}[]
\begin{minipage}{.3\linewidth}
\centering
\subfigure[]{\label{mach:a}\includegraphics[scale=.25]{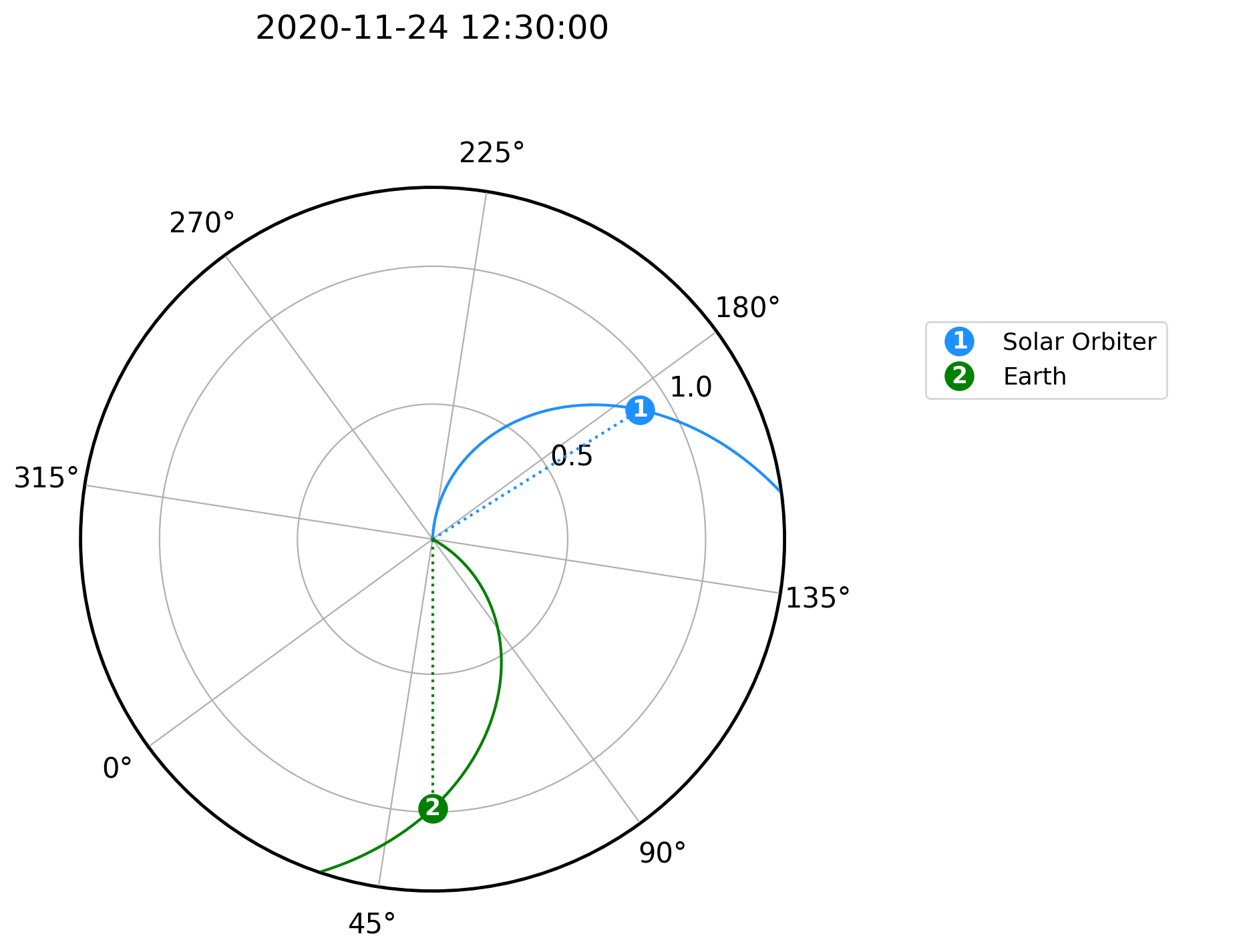}}
\end{minipage}%
\begin{minipage}{.3\linewidth}
\centering
\subfigure[]{\label{mach:b}\includegraphics[scale=.25]{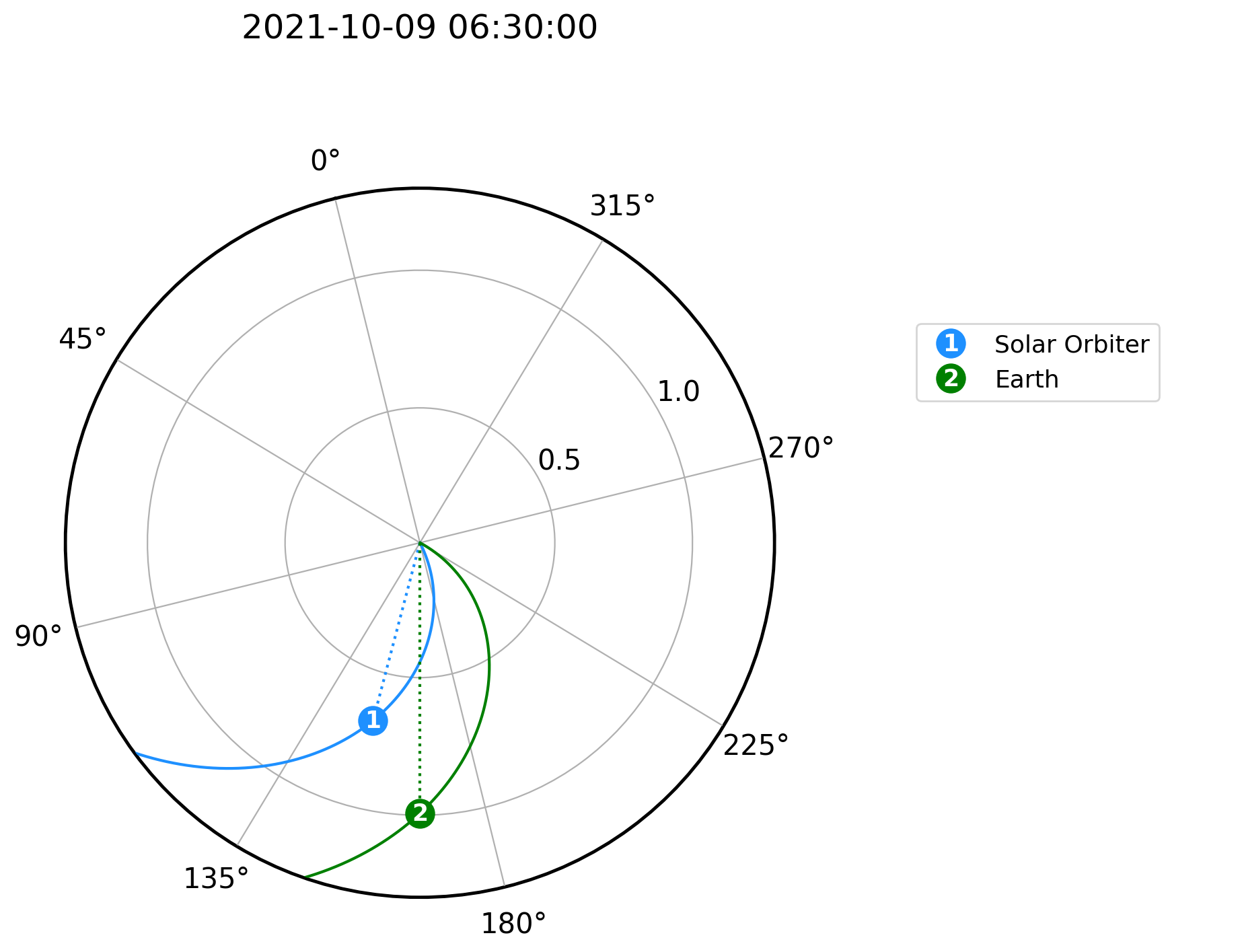}}
\end{minipage}
\centering
\begin{minipage}{.3\linewidth}
\centering
\subfigure[]{\label{mach:c}\includegraphics[scale=.25]{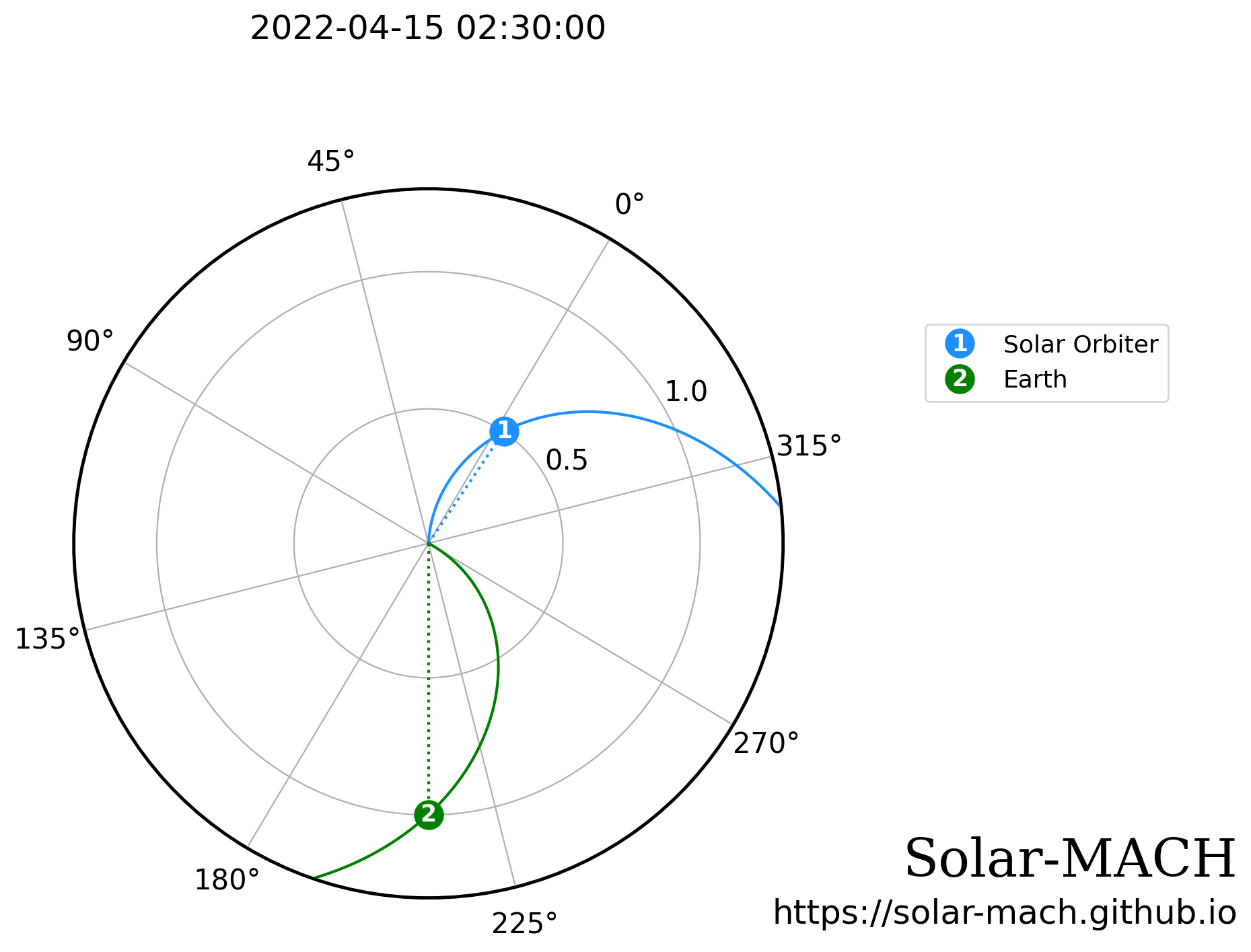}}
\end{minipage}
\caption{Position of \textit{Solar Orbiter} on 2020 November 24 (a), 2021 October 9 (b), and 2022 April 15 (c). The grid in black corresponds to the Carrington coordinate system. These polar plots are generated using the \href{https://serpentine-h2020.eu/tools/; Gieseler et al. 2022).}{Solar-MACH tool} \citep{Gieseler2023}.}

\label{solarmach}
\end{figure*}

\begin{figure*}
    \centering
    \includegraphics[width=\textwidth]{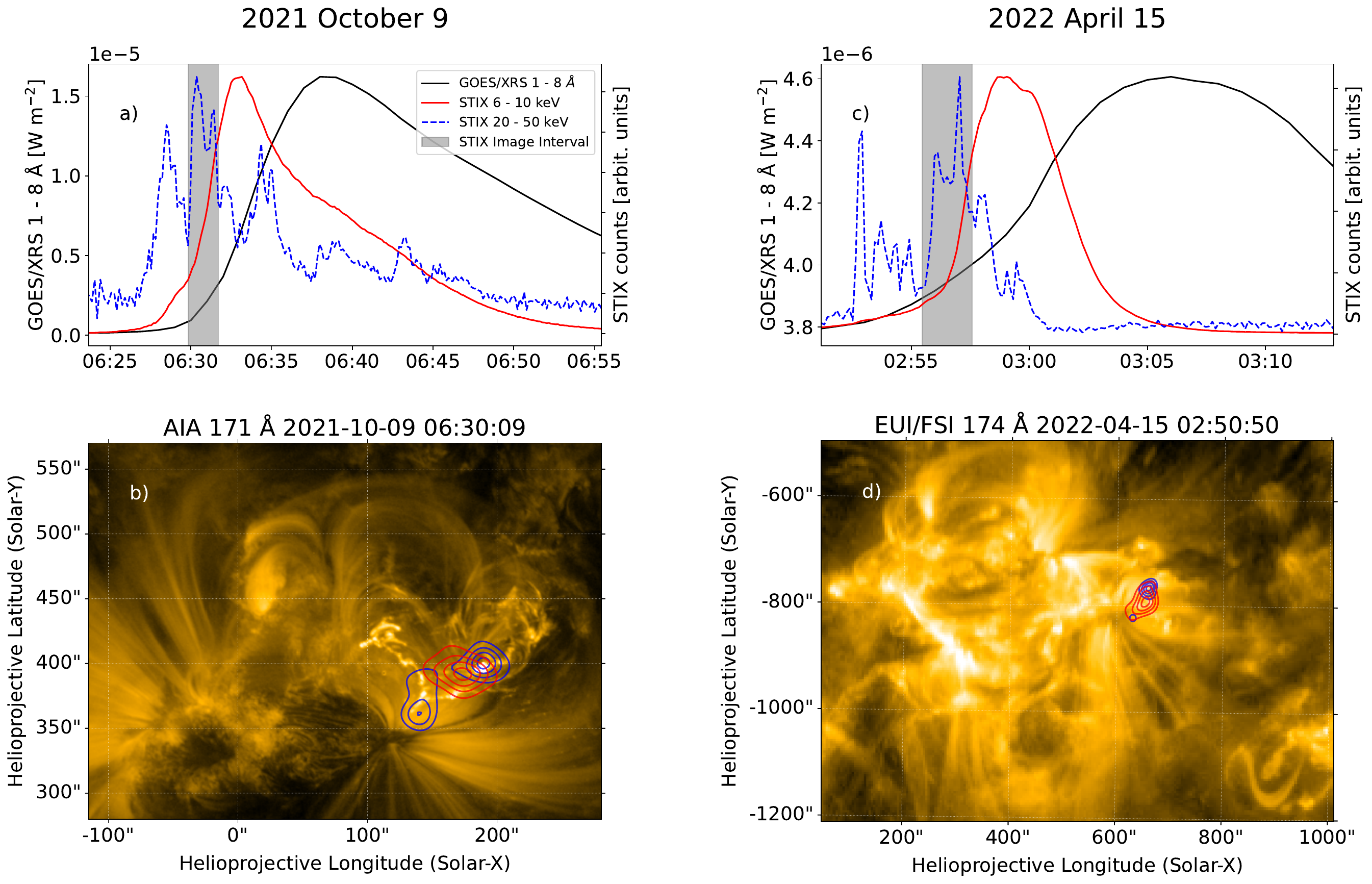}
    \caption{X-ray and EUV observations of the 2021 October 9 (left panels, a \& b) and 2022 April 15 (right panels, c \& d) events. {\it Top panels (a \& c)}: X-ray lightcurves from the \textit{GOES}/XRS 1\,--\,8\,\AA\ channel (black), and the STIX 6\,--\,10\,keV (red, thermal) and 20\,--\,50\,keV (blue dashed, non-thermal) spectral ranges. The grey regions show the intervals over which STIX counts were integrated to produce images. {\it Bottom panels (b \& d)}: STIX X-ray image contours overlaid on closest available EUV images. Panel~b shows an AIA~171\,\AA\ image reprojected to \textit{Solar Orbiter}'s viewing position. Panel~d shows an EUI Full Sun Imager 174\,\AA\ image. The STIX contour colors correspond the same spectral ranges shown in the top panels and the contour levels correspond to 30\%, 50\%, 70\% and 90\% of the maximum intensity in each spectral range. The STIX images were produced with the CLEAN algorithm using a CLEAN beam of 20''. The pointing uncertainty of these STIX observations is 10''.}
    \label{fig:euv_xray}
\end{figure*}

We select 43 events observed by EPD between July 2020 and April 2022 which shows clear velocity dispersion, reaching at least 50\,keV energies, and are associated with type III solar radio bursts. The selection criteria of an enhancement in electron fluxes at energies above 50\,keV is used to maximise the chance of having associated plasma oscillations, based on predictions of simulation work showing below a certain energy, the level of Langmuir wave being locally generated is very low \citep{Lorfing2023}. Of the 43 events selected, only 10 have both associated Langmuir waves and radio emission, as well as an electron flux significantly higher than the background flux measured by the EPD detectors STEP and EPT. We choose three example events (2020 November 24, 2021 October 9, and 2022 April 15) that each present different spectral behaviours. For each event, the position of \textit{Solar Orbiter} and the Earth are shown in Figure \ref{solarmach}. 
Figure~\ref{fig:euv_xray} shows EUV and X-ray observations, providing insight into the the active region geometry and the chromospheric impact locations (footpoints) of accelerated electrons for the 2021 October 9 and 2022 April 15 events. Such observations are not available for the 2020 November 24 event as the event occurred during the \textit{Solar Orbiter} cruise phase where the remote sensing instruments were not online, and the event likely originated on the non-Earth facing side of the Sun.

\section{Solar electron beams, Langmuir waves, and Type III solar radio bursts}

\begin{figure*}
\includegraphics[width=\textwidth]{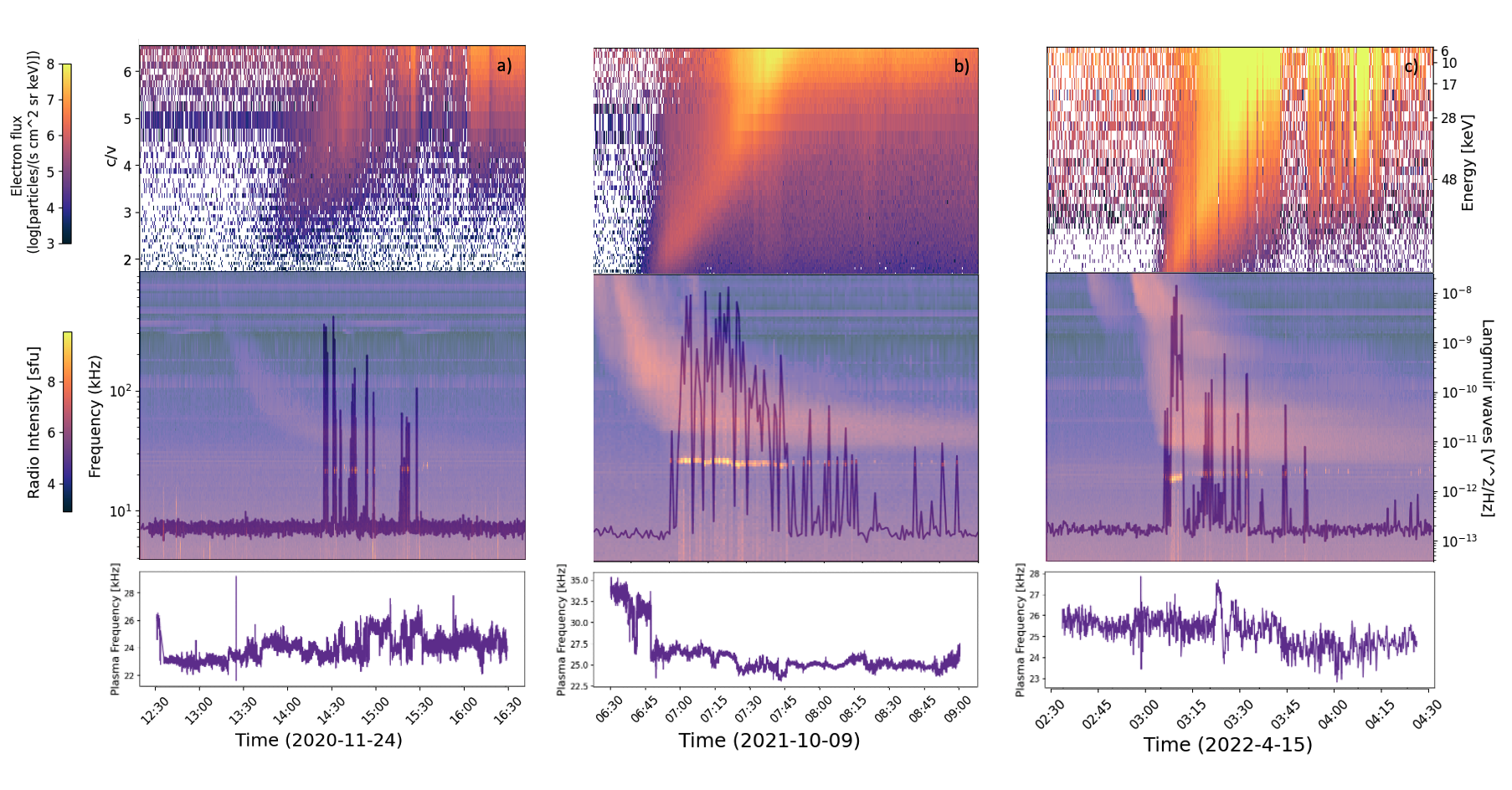}

\caption{\textit{Top}: Electron $c/v$ versus time plot (EPD-STEP). \textit{Middle}: Dynamic spectrum of the type III solar radio burst (RPW), and Langmuir wave [V$^{2}$/Hz] flux (RPW) above thermal level versus time both associated with the event shown on the top panel. \textit{Bottom}: electron plasma frequency $f_{\rm pe}$ [kHz] versus time. Column a) shows the 2020 November 24 electron event, column b) shows the 2021 October 9 electron event, and column c) shows the 2022 April 15 electron event. In the three cases \textit{Solar Orbiter} was at 0.901, 0.679, and 0.504\,au respectively.}
\label{gomezforpaper}
\end{figure*}

\textit{Solar Orbiter} observes solar electron beam events which may be associated with Langmuir wave events and type III solar radio emission. Three events displaying a combination of all three phenomena are shown on Figure \ref{gomezforpaper}, simultaneously measured by the spacecraft. This figure is composed of 3 subfigures: Figures \ref{gomezforpaper} a) (2020 November 24 at 0.901\,au), \ref{gomezforpaper} b) (2021 October 9 event at 0.679\,au), and \ref{gomezforpaper} c) (2022 April 15 at 0.504\,au). Each subfigure is comprised of three panels. The top panel of each figure shows the electron beam via its electron flux. For each event, we use the STEP electron fluxes in the 2-475\,keV energy range. In order to identify one (or more) electron beams, the electron spectrograms are analysed. The start ($t_{0}$) and end ($t_{1}$) times are identified using the observed enhancements of energetic electron fluxes. We manually set the time bounds of the electron distribution function plot to be 30 mins before and after visible flux enhancement. 

On the $y$-axis of each top panel, $c/v$ is displayed, where $v$ is the electron speed obtained from the energy at the centre of each bin, converted to speed in the spacecraft reference frame. Therefore, in the top panels for Figures \ref{gomezforpaper} a), \ref{gomezforpaper} b), and \ref{gomezforpaper} c), particle speed increases from top to bottom, as routinely done in studies of Solar Energetic Particle events \citep[e.g.,][]{Dresing2023}.

The middle panel of each figure shows the dynamic spectrum of the associated type III solar radio burst (4-900\,kHz), and overplotted, the Langmuir wave flux above thermal level. Figure \ref{gomezforpaper} a) shows locally generated Langmuir waves 5 orders of magnitude above the background solar wind plasma. We use an estimate of the plasma frequency, obtained semi-automatically from the quasi-thermal noise spectra, to integrate the spectral power in a broad band around the plasma frequency \citep[e.g.,][]{Robinson1993,Ergun1998,Vidojevic2010}. The Langmuir waves are also visible as yellow patches on the dynamic spectrum just below the type III solar radio burst. The Langmuir waves appear with the beam at energies around $c/v$ = 3.5 (50\,keV) at 14:25UT. Figure \ref{gomezforpaper} b) shows the 2021 October 9 electron event at 0.679\,au with an associated type III solar radio burst. We observe Langmuir waves locally generated from 07:00~UT, at 5 orders of magnitude above thermal level, co-temporal with the $c/v$ = 2.1 (75\,keV) electrons. Figure \ref{gomezforpaper} c) shows the 2022 April 15 electron event at 0.504\,au, the associated Langmuir waves, and the subsequent emission in the radio spectrum. The locally generated plasma wave, 5 orders of magnitude above the solar wind plasma, seems to be co-temporal with the electron beam at $c/v$ = 2 (80\,keV) electrons at 03:10UT.

The bottom panel of each figure shows the electron plasma frequency $f_{\rm pe}$ as a function of time for the event. Figures \ref{gomezforpaper} c), b), and a) in order of increasing heliospheric distance which would imply a decreasing electron density and therefore a decreasing plasma frequency which is in fact not the case because the time variations in the electron density dominate over the general radial variations for the periods analysed in this work. 

\

\section{Solar Electron Beam Spectra}
\label{spectra}

For this study, we use Level 2 products of SWA/EAS observations. We specifically analyse differential directional flux measurements. We first average the obtained flux, over the elevation and azimuth directions that are within the field of view of STEP instrument. In order to reduce the large measurement errors that are associated with the small number of counts obtained in the high energy range of SWA/EAS, we also average the flux over 30 consecutive measurements. We then detect the peak of the flux in each individual energy step and within the time intervals of the events we study. 

The peak electron flux during each event as a function of energy from EPD/STEP-EPT electron data is obtained by looping over each energy channel to obtain a time series of the electron flux at each energy and selecting the maximum of each trace between $t_{0}$ and $t_{1}$. The median value of the flux over this 30 minute window is also obtained for each energy channel.

\begin{figure*}
\centering
\begin{minipage}{.32\linewidth}
\centering
\subfigure{\label{peak:a}\includegraphics[width=5.5cm]{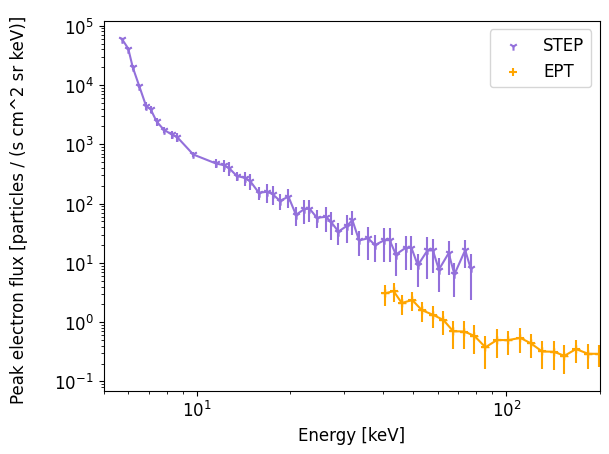}}
\end{minipage}%
\begin{minipage}{.32\linewidth}
\centering
\subfigure{\label{peak:b}\includegraphics[width=5.5cm]{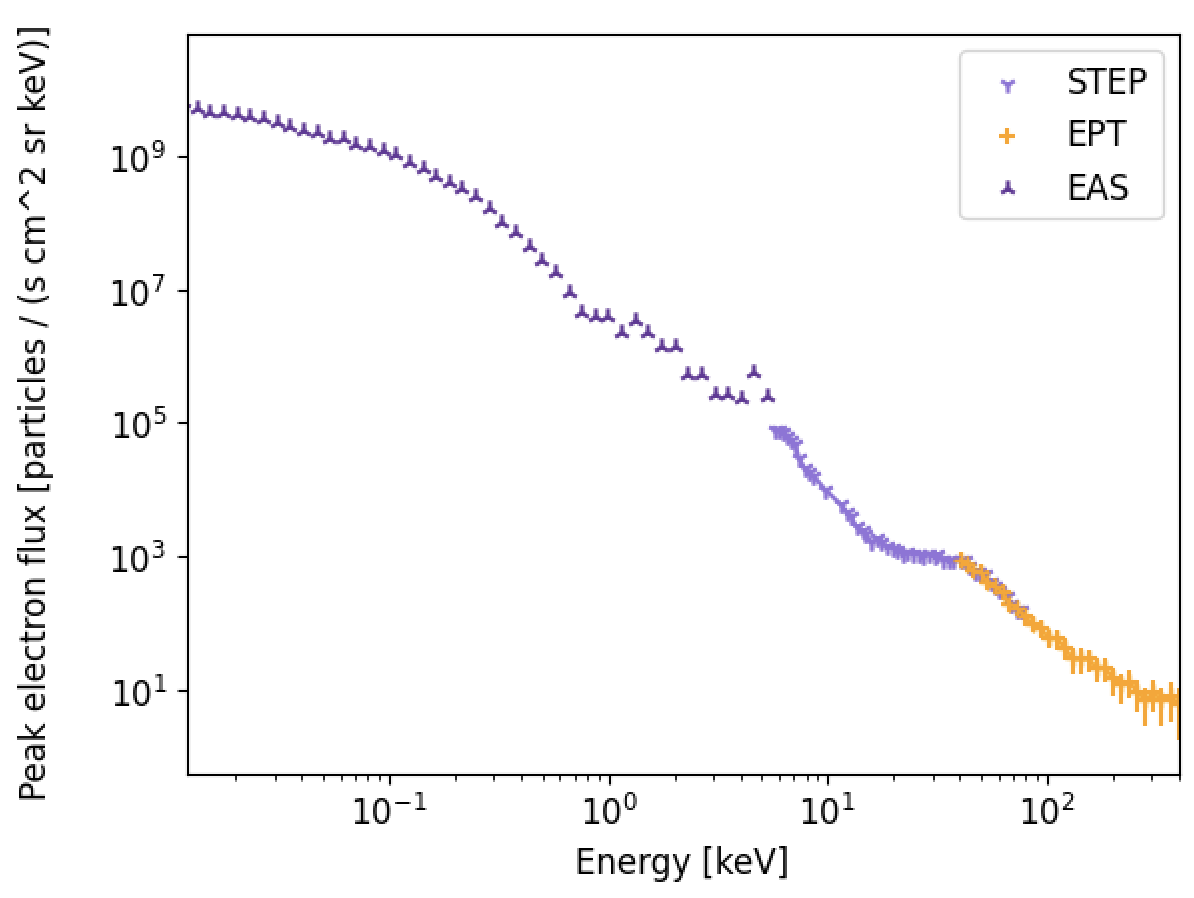}}
\end{minipage} 
\centering
\begin{minipage}{.32\linewidth}
\subfigure{\label{peak:c}\includegraphics[width=5.5cm]{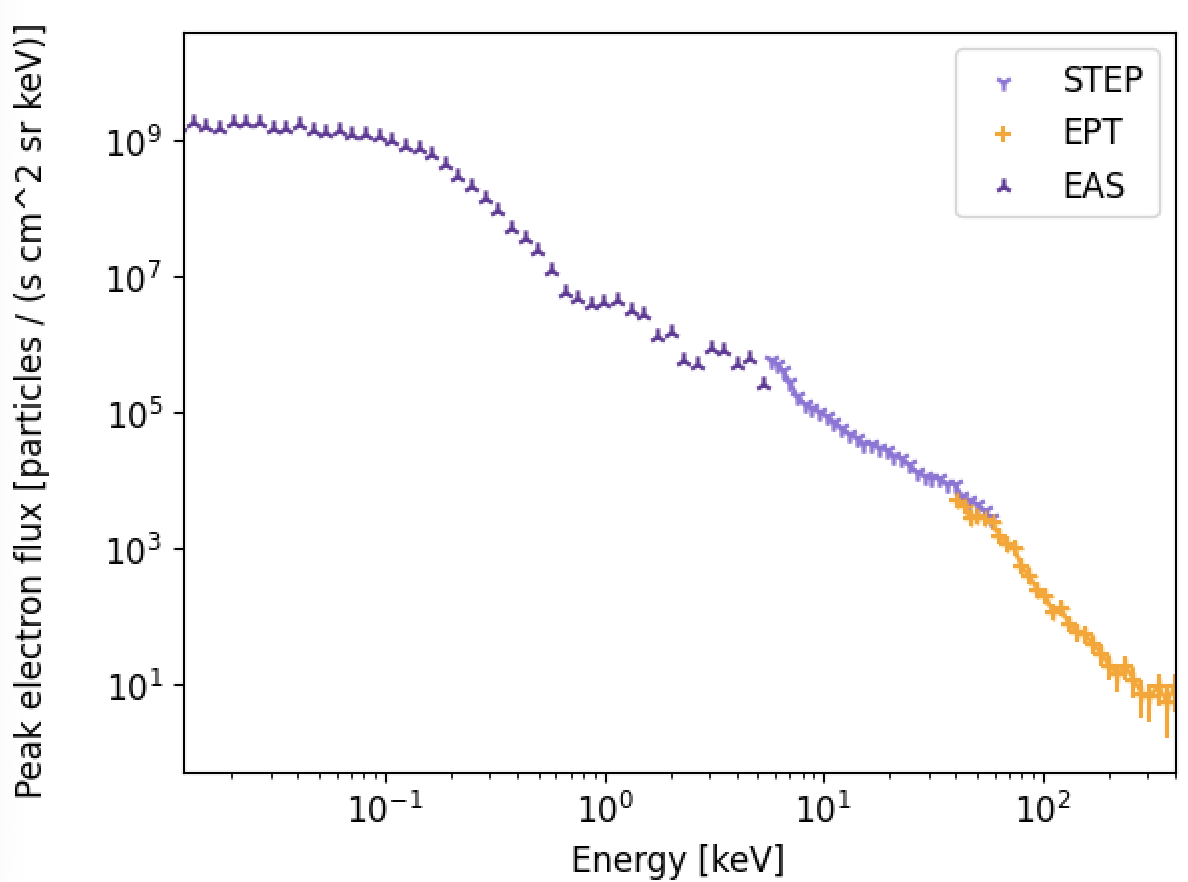}}
\end{minipage}

\caption{(a) Single power--law electron spectrum and associated uncertainties (STEP/EPT) for the 2020 November 24 electron event in figure \ref{gomezforpaper} a). (b) Triple power--law electron spectrum (EAS/STEP/EPT) for the 2021 October 9 electron event in Figure \ref{gomezforpaper} b). (c) Double power--law electron spectrum (EAS/STEP/EPT) for the 2022 April 15 electron event in Figure \ref{gomezforpaper} c).}
    \label{peakflux}
\end{figure*}

Errors on the measurements of the SWA electron peak flux calculated using different methods (e.g., standard deviation, Poisson error) translate into negligible percentage errors. These errors are too small to appear on Figure \ref{peakflux} despite the relevance of explicitly showing error bars on measurements in the energy range around the sewing of the SWA and EPD distributions. We use the errors on the EPD electron peak flux provided by STEP and EPT. Under the assumption of Poisson statistics for particle counting data, the errors in the particle counts during each accumulation time $N$ are assumed to be $\sqrt{N}$. This error is propagated to the intensities by dividing $N$ by the product of the geometric factor, the energy width of the channel, and the accumulation time. Close to the background levels, $N$ is small and the relative error can be large, but near the peak of big events, $N$ is large and the relative error becomes negligible. Errors in the geometric factor or the energy window or other possibles sources of systematic error are not considered because they are unknown. EPD error bars represent purely the statistical uncertainty in the counting rates. This error should be propagated quadratically if some re-averaging is performed \citep{EPD}.

The electron spectra as a function of energy for the 2020 November 24, 2021 October 9, and 2022 April 15 events can be seen in Figures \ref{peakflux} a), \ref{peakflux} b), and \ref{peakflux} c) respectively. SWA/EAS samples the electron range from 1 eV to 5\,keV, EPD/STEP samples the electron range from 5\,keV to 70\,keV, and EPD/EPT samples the electron range from 70 to 140\,keV between them covering a flux range of 10$^{3}$ to 10$^{12}$ particles/s cm$^{2}$ sr keV. We present the spectrum in the energy range 5\,eV-120\,keV while \textit{Solar Orbiter} capabilities can measure electrons from 1\,eV up to 30\,Mev. As detailed in Section \ref{EPD}, discrepancies in the EPD data between the STEP and EPT units after March/April 2021 required some adjustments in the electron flux measured by EPT. SWA/EAS data was not available for the 2020 November 24 event due to an instrument switch off.

Figure \ref{peakflux} a) shows a single power--law spectrum (STEP-EPT) with no inflexion points for the 2020 November 24 event at 0.901 AU. Figure \ref{peakflux} b) features multiple spectral breaks in the EAS and STEP data for the 2021 October 9 event at 0.679\,au. Some of these features are referred to as 'knees' - a spectrum with a low spectral index followed by a spectrum with a higher spectral index - and as 'ankles' - a spectrum with a high spectral index following by a spectrum with lower spectral index. The EAS electron spectrum shows an ankle at 700\,eV. At the junction between the EAS and STEP electron data around 5\,keV, we observe a jump. The STEP electron spectrum displays an ankle at 13\,keV followed by a knee at 35\,keV. 

Figure \ref{peakflux} c) shows several spectral breaks for the 2022 April 15 event (0.504\,au). We observe a knee in the STEP electron spectrum at 50 keV, preceded by an ankle at 8\,keV. A second knee, a energies just above the junction between the EAS and STEP electrons is observed at 6\,keV, also preceded by an ankle in the EAS electron spectrum at 600\,eV. A third knee is visible at the lowest EAS electron energies around 40\,eV.

Between the knees in the higher deca-keV range and the ankle around 10 keV on Figures \ref{peakflux} b) and c), there is a flattening of the electron peak flux and evidence of diffusion in energy space. Both figures also show similar features in the EAS energy range.\\

\section{Observed Spectrum Evolution}

\begin{figure*}
\centering
\includegraphics[width=\textwidth]{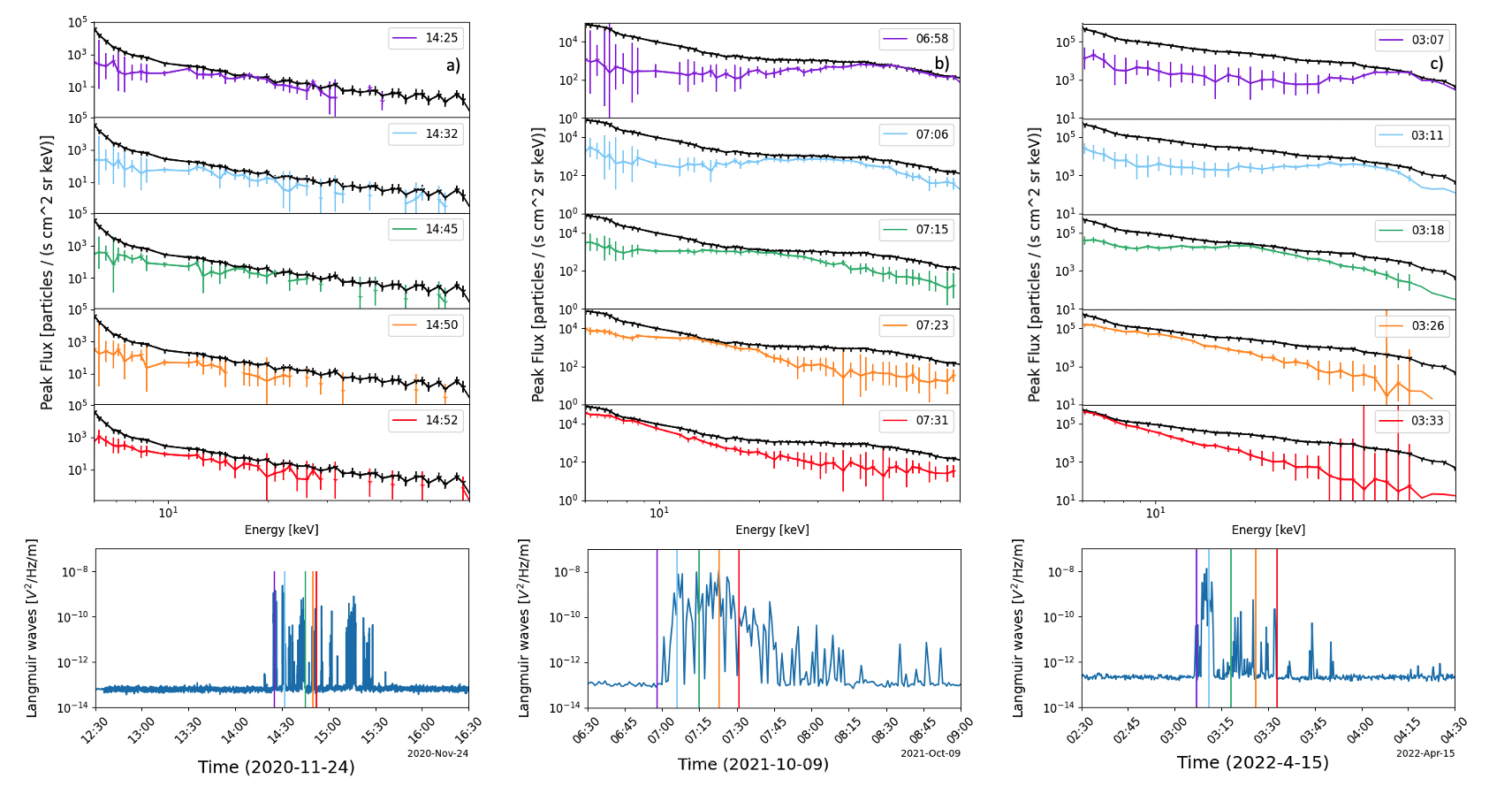}

\caption{Top panels of each column: Black dotted lines are the STEP peak electron flux observed during each event at each energy [keV]. Coloured lines are the electron velocity distribution function at the given times. Bottom panels of each column show the associated Langmuir wave flux with coloured lines indicating the timestamps at which the electron flux slices in the top panels were taken. Column a) shows the 2020 November 24 event, column b) shows the 2021 October 9 event, and column c) shows the 2022 April 15 event. 
   }
\label{vdfevol}
\end{figure*}

To understand the features that appear in the combined EAS-STEP-EPT electron spectra, we analyse the time evolution of the electron flux, as shown in Figures \ref{vdfevol} a), \ref{vdfevol} b), and  \ref{vdfevol} c). For each of the 2020 November 24, 2021 October 9, and 2022 April 15 events we plot the peak electron spectrum from Figures \ref{peakflux} a), \ref{peakflux} b), and \ref{peakflux} c) (black dotted line) on all panels of each figure. We then overplot, in colour, the instantaneous electron flux as a function of energy from Figures \ref{gomezforpaper} a), \ref{gomezforpaper} b), and \ref{gomezforpaper} c) at the different timestamps 1\,min that correspond to the times at which we observe Langmuir wave growth. The Langmuir wave flux is shown on the bottom panel of Figures \ref{vdfevol} a), \ref{vdfevol} b), and \ref{vdfevol} c). The vertical coloured lines overplotted on the Langmuir wave flux help identify the timestamps at which the electron flux as a function of energy curves were obtained. 

\section{Pitch Angle Distributions}
\label{PAD_sect}

\begin{figure*}

\includegraphics[width=\textwidth]{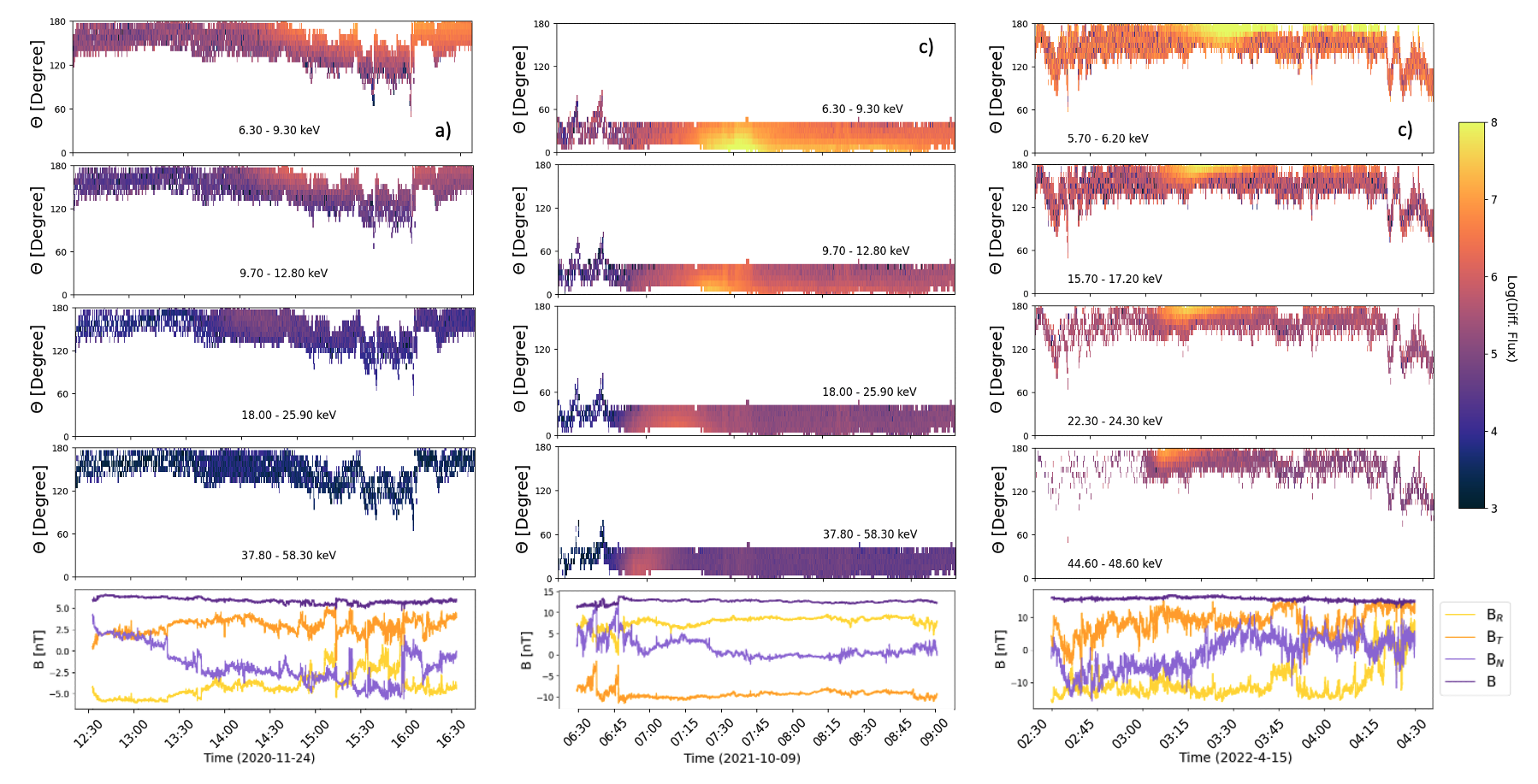}

\caption{Top panel of each column: Pitch angle ($\Theta$) distribution (EPD/STEP). Bottom panel of each column: Directional and total magnetic field (MAG). Column a) shows the 2020 November 24 electron event at 0.901\,au, column b) shows the 2021 October 9 electron event at 0.679\,au, and column c) shows the 2022 April 15 electron event at 0.504\,au.}

\label{pads}
\end{figure*}

Figures \ref{pads} (a), (b), and (c) show the pitch angle distributions (PAD) from STEP and magnetic field from MAG for the 2020 November 24, 2021 October 9, and 2022 April 15 events. As explained in section \ref{EPD}, Due to the limited STEP field of view, white areas in the PAD spectrograms of Figure~\ref{pads} are outside of the sensor coverage. In all three cases, the magnetic field is relatively stable and the electron beams are field-aligned, inducing a reasonable angular coverage of the beams. Figure \ref{pads} a), \ref{pads} b) and \ref{pads} c) all show highly anisotropic beams. This is less clear on Figure \ref{pads} a), as the field-aligned part of the distribution is not well-covered by STEP for the event, and the energetic particle enhancement observed throughout the event is weaker than the other two cases presented. The 2020 November 24 event is less optimal because the electron beam is weaker and the data collected from 16:00UT onwards is poor. This may cause some artifacts in the lower energy part of the spectrum. 
We note that the 2020 November 24 event (Figure \ref{pads} a)) and the 2022 April 15 event (Figure \ref{pads} c)) occur in an inward polarity interval, with a peak near a pitch angle of 180\,$^{\circ}$, while the 2021 October 9 event (Figure \ref{pads} b)) occurs during an outward polarity interval, with a peak near a pitch angle of 0\,$^{\circ}$. On the bottom panel of Figure \ref{pads} c), we notice a variation in the B field that corresponds to a jump in the PA coverage on the top panel of the same Figure. 

\begin{figure*}
\begin{minipage}{\linewidth}
\centering
\subfigure{\label{fig:eptpadsa}\includegraphics[width=11cm]{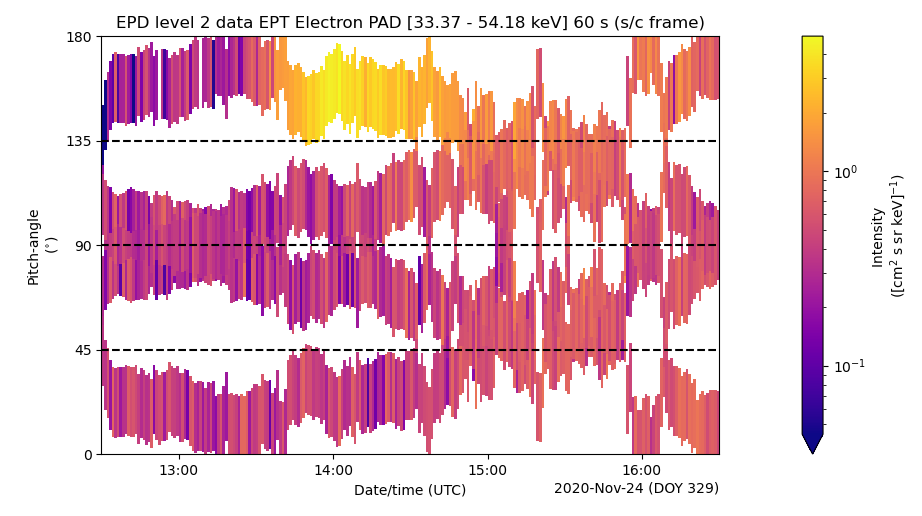}}

\end{minipage}%
\begin{minipage}{\linewidth}
\centering
\subfigure{\label{fig:eptpadsb}\includegraphics[width=11cm]{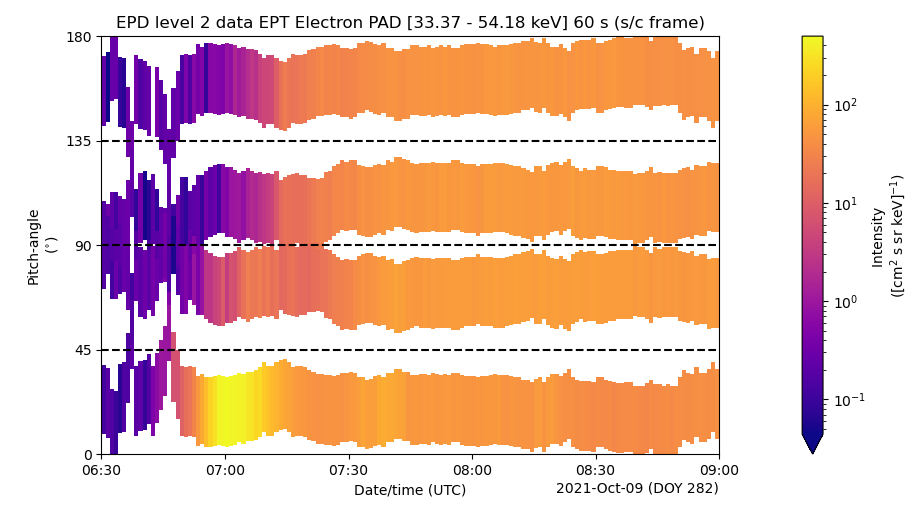}}
\end{minipage}%
\begin{minipage}{\linewidth}
\centering
\subfigure{\label{fig:eptpadsc}\includegraphics[width=11cm]{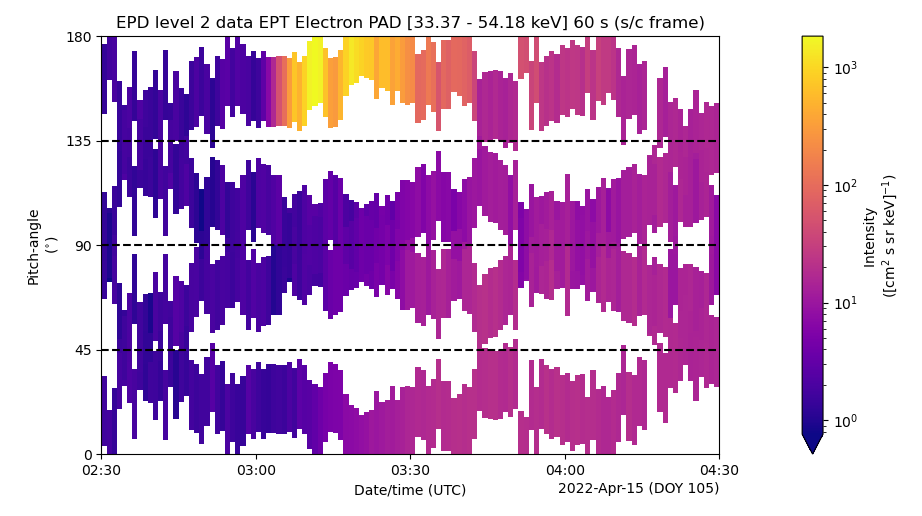}}

\end{minipage}%
\caption{Pitch angle distributions of 33.4-54.2\,keV electrons observed by EPT's four sensors during the 2020 November 24 (a), 2021 October 9 (b), and 2022 April 15 (c) electron events.}

\end{figure*}

Figures \ref{fig:eptpadsa}, \ref{fig:eptpadsb} and \ref{fig:eptpadsc} show the color-coded directional pitch-angle distribution of 33.4-54.2 keV electron intensities observed by EPT during the events on November 24, 2020, October 9, 2021 and April 15, 2022, respectively. 
While EPT provides only four looking directions, they normally offer better coverage of larger pitch-angles compared with STEP. As shown previously for STEP, the three events show pronounced anisotropy during the rising phase and the early decay phase, with higher fluxes observed by the EPT telescope pointing towards the Sun along the nominal Parker spiral direction. The first event was weak and the fluxes recovered the background level less than three hours after the onset. Conversely, the 2021 October 9 and 2022 April 15 events show strong increases and a long-lasting, nearly isotropic decay phase, which is typically observed during Solar Energetic Particle (SEP) events when the scattered particle population becomes dominant. 

\section{Discussion}
\label{discussion}

We observe solar accelerated electron events at different distances from the Sun, with associated Langmuir waves, and type III solar radio bursts as seen on Figures \ref{gomezforpaper} a), \ref{gomezforpaper} b), and \ref{gomezforpaper} c). These events are well connected and the spacecraft intercepts the central area of the electron beam in the interplanetary medium. The electron spectra combining EAS, STEP, and EPT electrons between 1\,eV and 300\,keV in Figure \ref{peakflux} obtained from the events in Figure \ref{gomezforpaper} show different spectral features such as knees and ankles, whose origin is yet to be understood. 

\subsection{The 2021 October 9 Event}
\subsubsection{Spectral break: the knee}

Observations at 1\,au show a spectral break in the deca-keV range \citep[e.g.][]{Krucker2009} that corresponds to a knee. \citet{Krucker2009} found the average energy this break occurred at was around 60\,keV using WIND/WAVES data. Simulations  comparing results from a free-streaming electron beam propagating without interactions to those with a beam resonantly interacting with the plasma has shown that the interactions could cause spectral breaks to arise in the electron spectrum \citep{Kontar2009,Reid:2010aa,Reid2013}, with a steepening in the spectrum at higher energies, above 60\,keV. 

A further look into the evolution of the observed electron flux and Langmuir waves for the 2021 October 9 event (0.679\,au), in Figure \ref{vdfevol} b), shows that the flux of higher energy electrons (80\,keV) that arrive at the spacecraft first are not high enough to produce a significant level of Langmuir waves, and we do not observe any Langmuir wave flux enhancement. At 06:58UT we observe a bump in the electron flux around 60\,keV, showing a positive velocity gradient $\partial f/\partial v > 0$. This condition is necessary for the beam to be unstable to Langmuir wave growth, yet there are no Langmuir waves observed at this point in time (bottom panel in Figure \ref{vdfevol} b)). Figure \ref{gomezforpaper} b) shows that the 50 keV electrons arrive with the onset of Langmuir waves activity at 07:00~UT. Velocity dispersion is seen as the bump moves down in velocity space as slower electrons arrive at the spacecraft. 
At 07:06UT, the distribution flattens between 25 and 50\,keV (knee in Figure \ref{peakflux} b)), due to quasilinear relaxation. The formation of the knee at 50 keV is due to a decrease in the peak flux of the electrons at any one point in time, related to the loss of energy from quasilinear relaxation due to Langmuir wave growth. The peak flux tracks the top right of the flux curve (Figure \ref{peakflux} b)) at any one point in time. As a result, the peak flux spectrum displays a lower power-law below 50\,keV, whose magnitude depends both on the level of Langmuir waves generated and the initial beam parameters \citep{Reid2013}. This is the first time these phenomena are observed in situ by virtue of the high temporal and spectral resolution of the \textit{Solar Orbiter} data. At the same time as the 50\,keV knee appears, we observe a Langmuir wave flux 5 orders of magnitude above the thermal level, highlighted as the vertical green line on the bottom panel of Figure \ref{vdfevol} b). The bottom panel of Figure \ref{gomezforpaper} b) shows a drop in the plasma frequency from 35 to 25\,kHz shortly after 06:50UT, just before we observe Langmuir waves. This appears to enhance the Langmuir wave intensity generated by the electron beam, with the Langmuir waves showing less clumpy behaviour in time. The lack of Langmuir wave clumps could be due to a reduction in the background electron density fluctuations \citep[e.g.][]{Reid:2017ab}.


\subsubsection{Spectral break: the ankle}
The beam-plasma interactions result in the electron flux diffusing down in energy space, forming a plateau. The lowest energy in the plateau decreases with time until it reaches the energies associated with the background solar wind plasma, and ceases to be able to diffuse any further. The peak flux is no longer able to be reduced as much any more due to the reduction of this quasilinear relaxation, and leads to the formation of an ankle in the peak flux spectrum. This is shows around 07:23~UT in Figure \ref{peakflux} b), where we can see the peak flux of electrons increasing at the energies lower than 13 keV. At 07:31UT the distribution is aligned with the peak electron flux and $\partial f/\partial v$ is now negative at all points above 5 keV. After 07:31UT, the Langmuir wave flux decreases as there less electrons that are unstable to Langmuir wave growth. 

At lower energies around 600 eV, there is a second ankle in the peak flux spectrum. This ankle is where the electron beam meets the strahl/halo in the background solar wind plasma.



\subsubsection{Pitch Angle}
The associated pitch angle distribution in Figure \ref{pads} (b) shows the beam is highly anisotropic as the flux enhancement is greatest near a pitch angle of zero degrees. There are no significant changes in the electron flux at energies at which we observe a break in the electron spectrum. Pitch angle scattering is energy dependent \citep{Droge2000} and can be modelled using a Fokker--Planck approach \citep[e.g.][]{Kontar2014}. A strong pitch angle diffusion coefficient can cause a flattening in the electron peak flux. However, we do not observe a significant change in the pitch angle for the energies associated with either the knee or the angle in our event. 

Two observational and simulation studies \citep{Strauss2020,Dresing2020} look at different shapes of electron energy spectra. They show that the spectral changes and breaks at energies above 100 keV and further away from the Sun ($R>$ 1\,au) are potentially caused by pitch angle scattering and they conclude that a spectral break around 60\,keV would be caused by Langmuir wave growth \citep{Krucker2009}. \citet{Dresing2020} further specifies they do not study low energy electrons. \citet{Droge2018} also show that low energy electrons suffer very little pitch angle scattering. Simulations \citep{Kontar2009,Reid2013,Reid2018} and observations are in agreement and show that within the deca-keV range, the wave-particle interactions cause the spectral break and thus explain the flattening of the electron flux for the 2021 October 9, as seen on Figure \ref{peakflux} b).

\subsection{The 2022 April 15 Event}

The 2022 April 15 event displays a spectral break at 50 keV that is a knee, similar to the spectrum of the 2021 October 9 event, as can be seen on Figures \ref{peakflux} c) and \ref{peakflux} b), respectively. The evolution of the electron flux follows the same trend where diffusion in energy space occurs at the same time that Langmuir waves are observed. The peak magnitude of the Langmuir waves is the same as the 2021 October 9 event, but they last for a much shorter duration, as can be seen by comparing the bottom panels of Figures \ref{vdfevol} c) and \ref{vdfevol} b), despite the electron flux in both events being comparable in magnitude. 

The change in the B field associated with the 2022 April 15 event shown on the bottom panel of Figure \ref{pads} c) between 03:40UT and 03:50UT causes an interruption in the STEP angular coverage of the pitch angle region near the centre of the beam, around 180$^{\circ}$ between these two timestamps. STEP therefore misses the centre of the beam despite it still being present because the sensor does not point in the correct direction with respect to the B field. This is visible as a drop in the electron flux in the same ten minute interval on the top panel of the Figures \ref{gomezforpaper} c). This does not seem to have affected the production of co-temporal Langmuir waves. It is equally possible for true flux dropouts to happen if the observer temporarily shifts to a neighbouring flux tube, empty of energetic particles. The intensification in the RPW spectrogram on the middle panel of Figure \ref{gomezforpaper} c) appears at lower frequencies than for the previous two events despite its shorter heliocentric distance (0.504\,au). The locally measured electron plasma frequency $f_{\rm pe}$ (bottom panel of Figure \ref{gomezforpaper} c)) varies between 24 and 26\,kHz. $f_{\rm pe}$ relates directly to the electron density which is itself a function of heliocentric distance, highest closer to the Sun. We however observe that the plasma frequency is around 24 kHz at 0.901\,au and drops from 33\,kHz to 25\,kHz at 0.679\,au. Figure \ref{pads} c) shows that the sensor is aligned with the centre of the beam, around 180$^\circ$, measuring the real peak of the electron flux.

\subsection{The 2020 November 24 Event}

The electron flux measured by EPD for the 2020 November 24 event (top panel of Figure \ref{gomezforpaper} a)) is 2-3 orders of magnitude smaller than the flux for the other two events presented in this work. This event is fainter, and therefore the associated Langmuir wave flux is incidentally lower. The PAD for the 2020 November 24 seen on Figure \ref{pads} a) shows that the electron beam is B field aligned. The ability of RPW to observe the associated Langmuir oscillations and radio emission is however decreased by the phenomena's low intensity and $k$-vector with respect to the antenna directions. We observe that the peaks in the Langmuir wave flux for this event (middle panel of Figure \ref{gomezforpaper} b)) are much narrower than for the 2021 October 9 event or the 2022 April 15 event shown on the other two columns of the middle panel on the same figure. This translates into a single power--law spectrum as seen on Figure \ref{peakflux} a) and the absence of inflexion points. When looking at the evolution of the electron flux over the period of the event (Figure \ref{vdfevol} a)), the bump in the flux, and the subsequent velocity dispersion due to lower energy electrons arriving at the spacecraft are barely noticeable. The flattening caused by the quasilinear relaxation from the wave-particle interactions is equally poorly visible, but we notice some slight quasilinear diffusion around 14:30UT. However the 2021 October 9 event and the 2022 April 15 event have shown how the electron flux is modified by the wave particle interactions, and how this causes breaks to appear in the electron spectrum around 50 keV. We believe that this event might display a double or triple power--law in its spectrum if its centre is correctly observed in alignement with the EPD sensors.\\

All three of our events were detected at different distances from the Sun, with the 2022 April 15 event occurring at 0.504\,au whilst the 2021 October 9 event occurring slightly further away at 0.679\,au, but the electron spectra of both events remains similar (as can be seen on Figures \ref{peakflux} c) and \ref{peakflux} b)). 
A stronger solar eruptive event can lead to the beam travelling further away from the Sun and being detected more strongly than a weaker event close to the Sun. Simulations have shown that stronger events generate Langmuir waves at higher energies further away from the Sun \citep{Lorfing2023}. This opens prospects to further investigate the radial variations of these processes with wider sample of EPD events.




\section{Conclusion}

We looked at electron fluxes, electron spectra, pitch angle distributions, and associated Langmuir wave and type III solar radio bursts for 43 events between 2020 July 23 and 2022 April 15 detected \textit{in situ} by \textit{Solar Orbiter}. We found 10 of these events showed detectable electron fluxes in EPD/STEP and a signature of Langmuir waves in RPW. Using these events, we study for the first time at such a high temporal and spectral resolution, how resonant wave-particle interactions cause velocity dispersion and quasilinear relaxation in the electron flux, causing inflexion points to appear on the electron spectrum in the deca-keV range. We present three example events (2020 November 24, 2021 October 9, and 2022 April 15) at different distances from the Sun. Combining spectral data from EAS and EPT for the first time, the events show various spectral behaviours over the 5 order of magnitude in energy space. We show that the beam-plasma interactions result in a flattening of the electron flux at deca-keV energies where these interactions happen, and causes the appearance of a spectral break around 50\,keV in the shape of a knee. For one event, we find that around 15\,keV an ankle in the electron spectrum is formed. At these energies quasilinear relaxation decreases as the electron flux has less room to diffuse down in energy space, resulting in an increase in the peak flux spectrum at energies below 15 keV. Consequently, we attribute this ankle to be formed due to transport effects and not to the intrinsic shape of the injection spectrum at the Sun, as suggested by \citet{Lin1985}. In the EAS electron spectrum, another ankle is observed where the electron beam meets the strahl and halo of the solar wind plasma, around 600\,keV. We find that if the electron flux measured by \textit{Solar Orbiter} is weak, the associated electron spectrum displays a single power-law. For more intense events, spectral features like a double or triple power--law spectrum become visible, as previously observed \citep{Lin1982,Lin1985, Lin1990,Krucker2009,Dresing2021,Dresing2023,Wang2023}. Futhermore, we highlight the importance of using overlapping FOVs when merging data from different sensors and instruments like EAS1 (SWA) and STEP (EPD). Lastly, an analysis of pitch angle distributions shows that for highly anisotropic beams, non-thermal electrons in the deca-keV range are more affected by wave-particle interactions than by pitch angle scattering, in line with previous results \citep{Strauss2020,Dresing2020}. Whilst we do not find a link between our results and heliocentric distance, this project opens prospects to further investigate the radial variations of these processes with wider sample of EPD electron events.


\section{Acknowledgements}
Solar Orbiter is a space mission of international collaboration between ESA and NASA, operated by ESA. Solar Orbiter Solar Wind Analyser (SWA) data are derived from scientific sensors which have been designed and created, and are operated under funding provided in numerous contracts from the UK Space Agency (UKSA), the UK Science and Technology Facilities Council (STFC), the Agenzia Spaziale Italiana (ASI), the Centre National d’Etudes Spatiales (CNES, France), the Centre National de la Recherche Scientifique (CNRS, France), the Czech contribution to the ESA PRODEX programme and NASA. Solar Orbiter SWA work at UCL/MSSL was funded under STFC grants ST/T001356/1, ST/S000240/1, ST/X002152/1 and ST/W001004/1. Solar Orbiter EUI at UCL/MSSL was funded under STFC grants ST/P002463/1, ST/S00002X/1 and ST/T000317/1 \\

H. Reid, C.J. Owen and D. Verscharen acknowledge funding from the STFC Consolidated Grant ST/W001004/1.\\

C.Y. Lorfing and H. Reid acknowledge support from the Royal Society International Exchange Project IEC$\backslash$R2$\backslash$202175. \\

The UAH team acknowledges the financial support by the Spanish Ministerio de Ciencia, Innovación y Universidades FEDER/MCIU/AEI Projects ESP2017- 88436-R and PID2019-104863RB-I00/AEI/10.13039/501100011033.\\

This work has received funding from the European Unions Horizon
2020 research and innovation programme under grant agreement No.
101004159 (\href{ www.serpentine-h2020.eu}{SERPENTINE})\\

The RPW instrument has been
designed and funded by CNES, CNRS, the Paris Observatory, The Swedish
National Space Agency, ESA-PRODEX and all the participating institutes.\\

D.\ F.\ Ryan thanks Paolo Massa (Western Kentucky University) and Ewan Dickson (University of Graz) for their helpful clarifications.\\

\vspace{5mm}





\bibliography{bibliography}{}
\bibliographystyle{aasjournal}



\end{document}